# RadRotator: 3D Rotation of Radiographs with Diffusion Models

A Technical Report

Pouria Rouzrokh[1,2,§], Bardia Khosravi[1,2,§],
Shahriar Faghani[1], Kellen L. Mulford[2], Michael J. Taunton[2], Bradley J. Erickson[1],
Cody C. Wyles[2,*]

(1) Mayo Clinic Artificial Intelligence Laboratory, Mayo Clinic, MN, USA
(2) Orthopedic Surgery Artificial Intelligence Laboratory, Mayo Clinic, MN, USA

(§) Co-first authors (*) Corresponding author
Please email all correspondence to wyles.cody@mayo.edu

## Abstract

Transforming two-dimensional (2D) images into three-dimensional (3D) volumes is a well-known, yet challenging problem for the computer vision community. In the medical domain, a few previous studies attempted to convert two or more input radiographs into computed tomography (CT) volumes. Following their effort, we introduce a diffusion model-based technology that can rotate the anatomical content of any input radiograph in 3D space, potentially enabling the visualization of the entire anatomical content of the radiograph from any viewpoint in 3D. Similar to previous studies, we used CT volumes to create Digitally Reconstructed Radiographs (DRRs) as the training data for our model. However, we addressed two significant limitations encountered in previous studies: 1. We utilized conditional diffusion models with classifier-free guidance instead of Generative Adversarial Networks (GANs) to achieve higher mode coverage and improved output image quality, with the only trade-off being slower inference time, which is often less critical in medical applications; and 2. We demonstrated that the unreliable output of style transfer deep learning (DL) models, such as Cycle-GAN, to transfer the style of actual radiographs to DRRs could be replaced with a simple, yet effective training transformation that randomly changes the pixel intensity histograms of the input and ground-truth imaging data during training. This transformation makes the diffusion model agnostic to any distribution variations of the input data pixel intensity, enabling the reliable training of a DL model on input DRRs and applying the exact same model to conventional radiographs (or DRRs) during inference.

**Keywords**: Radiology, Radiograph, Digitally Reconstructed Radiograph, DRR, Artificial Intelligence, AI, Deep Learning, Machine Learning, Generative AI, Diffusion models, DDPM

**Website**: https://pouriarouzrokh.github.io/RadRotator
**Online Demo**: https://huggingface.co/spaces/Pouriarouzrokh/RadRotator

# 1 Introduction

Radiologic imaging plays a crucial role in the diagnosis and management of a wide range of musculoskeletal pathologies (1). Orthopedic surgeons frequently leverage imaging for various purposes, from preoperative planning to postoperative monitoring and assessment of complications (2). Even though three-dimensional (3D) modalities such as computed tomography (CT) and Magnetic Resonance Imaging (MRI) have their specific use cases in orthopedic applications, X-ray radiographs, including derivatives like intraoperative fluoroscopy, remain the most commonly chosen modality by orthopedic surgeons. Indeed, radiographs are so commonly obtained in orthopedic surgery that they are considered among routine examinations for surgeons (3).

Radiographs offer significant advantages over 3D imaging: they are more readily available, are cost-effective for patients and the health system, and expose patients to lower radiation levels. (4). However, radiographs have limitations. They often require multiple images from fixed angles to adequately visualize anatomical structures, a process that can be difficult to perform and introduces variability into the appearance of standard views. Likewise, in surgical and interventional settings, the dynamic capture of anatomical changes via fluoroscopy is essential but can be inefficient and heavily reliant on the expertise of physicians and radiographic technicians. Last but not least, 3D imaging offers distinct insights that cannot be obtained through single or multiple-view radiographs.

Choosing between plain radiographs and 3D imaging is thus often accompanied by some trade-offs in medical imaging. Yet, recent advancements in generative artificial intelligence (AI) may offer an optimal solution to medical practitioners. According to the Universal Approximation Theorem in machine learning sciences, deep learning (DL) models can approximate any continuous function with the desired degree of accuracy, given an appropriate network structure and sufficient training data. Leveraging this theory, it is feasible to develop a DL framework that can process one or more input plain radiographs and produce a radiographic view from any point, or even an entire 3D volume, accurately representing the anatomy present in the radiograph.

Several previous studies have provided preliminary evidence supporting the feasibility of such generative frameworks in medical domains. Ying et al. proposed a Generative Adversarial Network (GAN)-based (X2CT-GAN) method to create CT scans from two biplanar chest radiographs (5). They used a single-pass hybrid generator that converted the two radiographs directly into a 3D volume. Similarly, Ge et al. used a single-pass approach to create 3D cervical spine reconstructions from orthogonal neck radiographs (6). However, both approaches resulted in high reconstruction errors (> 3mm), which are not accurate enough for clinical use. Using a different approach, Shamanth et al. used multiple fluoroscopic images combined with a simple neural network to create 3D bone reconstructions of knees with 2 mm errors (7). However, in the domain of natural images, the current state-of-the-art 3D object generation relies on a "2D-diffusion uplifting" technique that creates multi-view two-dimensional (2D) generation and converts them to a 3D model (8).

In this report, we introduce a novel generative AI framework designed to use a single-view radiograph to generate radiographs from any point in 3D space and subsequently generate consistent virtual videos that visualize the patient's anatomy in three dimensions. Our work distinguishes itself from the previous efforts in several ways: 1. Instead of relying on GANs, our framework employs conditional Denoising Diffusion Probabilistic Models (DDPMs) (9), which are state-of-the-art in image generation, surpassing GANs in terms of image quality and diversity (10); 2. Our method requires only a single-view radiograph as a conditional input, contrasting with earlier studies that needed two orthogonal views for generating unseen views; 3. While we adhere to the convention of training our model on Digitally Reconstructed Radiographs (DRRs) derived from CT data, we also introduce a straightforward, yet effective training data transformation technique named *RandHistogramShift*. This technique ensures our model performs well on both DRRs and actual radiographs, eliminating the need for a separate style-transfer DL model, such as Cycle-GAN (11). This report is dedicated to outlining the technical specifics of our framework and providing both proof-of-concept and qualitative assessments of its performance on pelvis radiographs. We leave comprehensive and quantitative evaluations of our proposed pipeline for future work.

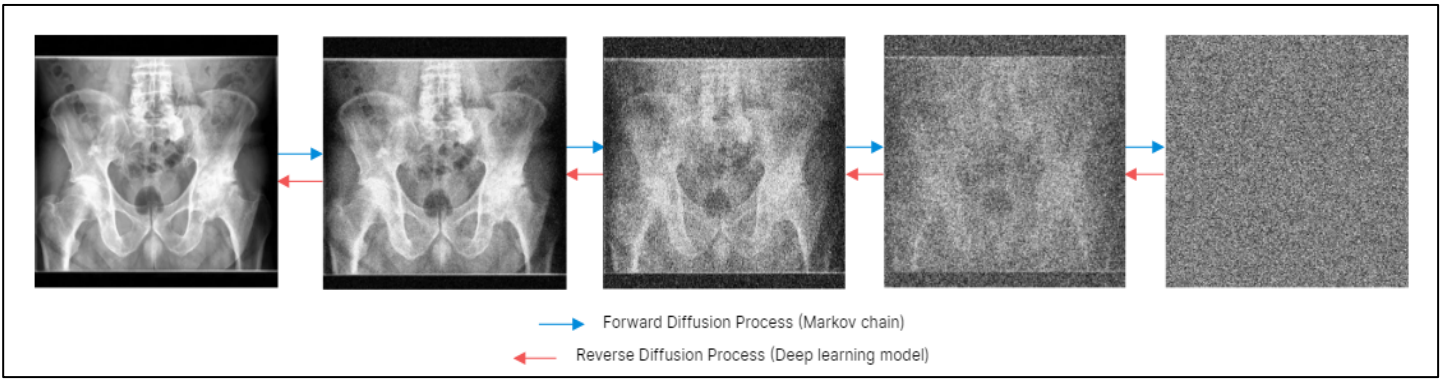

**Figure 1**. Overview of the forward and reverse diffusion process.

## 2 Preliminaries

### 2.1 Diffusion Models

Diffusion models are a recently introduced family of DL models in computer vision that learn to gradually convert random noise into a meaningful image mimicking the distribution of their training data (9). This approach to generating images is fundamentally different from the direct generation methods used by other types of models, such as GANs. Diffusion models work by first 'destroying' a copy of an image over a series of steps to turn it into random noise, and then learning to reverse this process to create new images from noise. This is achieved in two main phases: the forward pass and the backward pass (**Figure 1**).

In the forward pass, the diffusion algorithm incrementally adds Gaussian noise to an input image over a predetermined number of steps, until the image is completely transformed into random Gaussian noise. It is important to note that the added Gaussian noise at each step is sampled from distributions with pre-determined means and standard deviations, so it does not require training a network. The backward pass, which is where the learning happens, involves training a neural network to approximate the parameters of the distributions used for sampling random noise during the forward pass and reverse that pass step by step (i.e., 'denoising'). At each step, the model makes a prediction about what the less noisy image looked like one step before, starting from pure noise and gradually removing noise until it arrives at a meaningful image. This reverse process is guided by minimizing the difference between the ground truth less noisy images and the generated ones, or by using other loss functions that similarly optimize the denoising capabilities of the model.

The DDPM is a specific type of diffusion model that formalizes the diffusion process as a Markov chain and uses variational inference for the denoising process. The utilization of DDPMs in our framework represents a strategic adaptation to the unique requirements of medical imaging within the challenging framework of the generative AI trilemma; **Figure 2**. This trilemma encapsulates the struggle to simultaneously fulfill three key requirements of generative AI models: high sample quality, sample diversity, and fast sampling (12). DDPMs, distinguished by their ability to produce images of high quality with impressive mode coverage, are better alternatives to GANs and variational autoencoders (VAEs) (10). Their primary drawback, however, is the slower inference speed. In the context of medical imaging, where the accuracy and detail of images are paramount, the slower speed of DDPMs is often a reasonable trade-off.

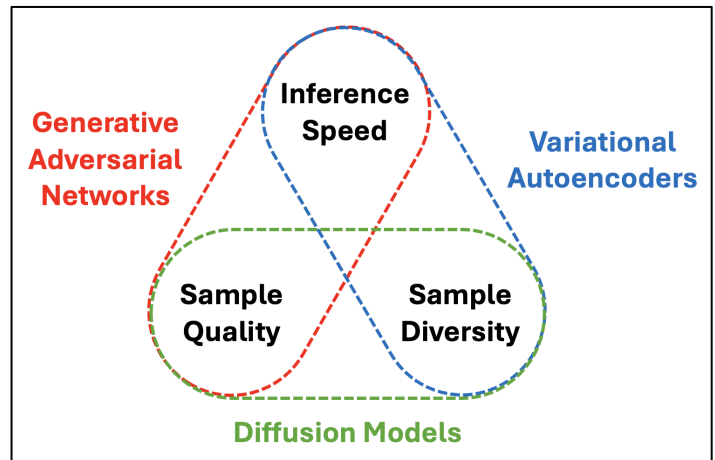

**Figure 2.** The generative artificial intelligence trilemma.

## 2.2 Digitally Reconstructed Radiographs (DRRs)

DRRs are a form of medical imaging that serves as a virtual simulation of traditional x-ray images, generated from three-dimensional CT scans (13). This means that unlike conventional radiographs, which capture a single 2D view of X-ray absorption through the body, DRRs are generated by simulating the radiograph projection through the 3D CT volume. To create each pixel in a DRR, an algorithm needs to estimate and sum the linear attenuation coefficients of the tissues that a simulated X-ray should pass through when traveling from a virtual X-ray source to the simulated 'film' or receptor. The resulting 2D image resembles a conventional radiograph, albeit sometimes with notable differences in brightness and contrast (**Figure 3**).

DRRs have several important applications in medical imaging and treatment planning (14,15). They can be used to compare CT data to conventional radiographs to ensure the treatment beams are properly aligned, such as in radiotherapy planning. DRRs can also be used for training and simulation in radiology and radiography, as they provide realistic X-ray images without exposing patients to radiation beyond the original CT. Additionally, DRRs have been used to aid in the identification of deceased individuals by comparing them to historical radiographic records. In the world of computer vision, DRRs have been shown to have another advantage as well: they can be used to train DL models to transform input radiographs into 3D reconstructions (5,16). In an ideal scenario, training such generative models requires access to many radiographs acquired from the same patient at the same time. Obviously, meeting such a requirement is not feasible. As a result, data scientists leverage CT data to generate numerous DRRs and subsequently train their models on these virtual radiographs.

However, the above approach has an important limitation. As the style of DRRs differs from conventional radiographs, all studies that aim to use DRRs for creating (or augmenting) training data for their models must find a solution to transfer the style of DRRs to that of typical radiographs before initiating the training. The most frequently explored category of solutions for this bottleneck is style transfer DL models such as Cycle-GAN. For example, Gao et al. have used DRRs generated from pelvic CT images to augment their training dataset for intraoperative landmark registration (17). Similarly, Loÿen et al. have used DRRs to create multiple views from CTs for patient-specific 3D modeling (18). To align the style of generated DRRs with original radiographs, both studies used a Cycle-GAN model for unpaired image-to-image translation (11). Despite the novelty of this approach, the downsides of using style transfer models for converting DRRs to radiographs are significant (19). Not only is training a separate DL model alongside the main model of interest costly and time-consuming, but the performance of style transfer models is not consistently appropriate. Style transfer models are inevitably trained on unpaired image-to-image datasets (i.e., datasets that contain images from two different styles that are not matched with each other), and, therefore, there is always a risk that the model cannot accurately discern the style and content of the source image. This leads to the potential issue of altering the content of a DRR (i.e., the patient's anatomy) during the process of transferring its style to that of a radiograph.

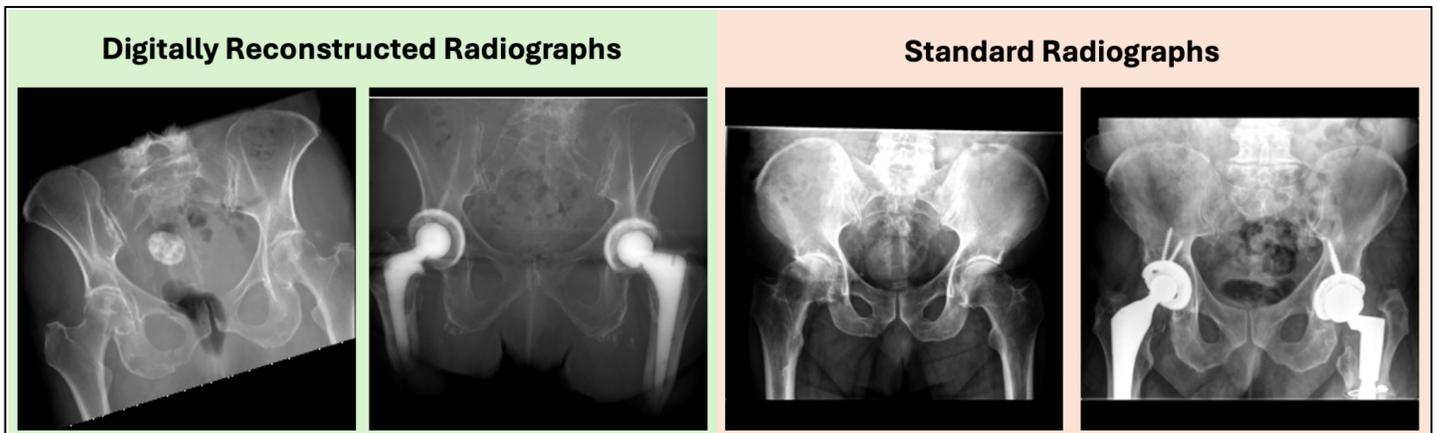

**Figure 3**. Digitally Reconstructed Radiographs (DRRs) vs. standard (conventional) radiographs.

## 3 Methods

In this report, we describe our proposed approach for rotating radiographs in 3D through the introduction of a DL pipeline that can receive a single AP pelvis radiograph (with or without hip arthroplasty prosthesis present in the image) and, as a proof-of-concept, rotate it up to 15 degrees in the x, y, and/or z planes. Theoretically, the same pipeline could be applied to any amount of rotation and any kind of input radiographs.

### 3.1 Data

After obtaining Institutional Review Board approval, we collected a total of 9,044 non-contrast-enhanced hip and pelvis CT studies acquired at three different Mayo Clinic sites (Minnesota, Florida, and Arizona) between 2000 and 2018. This cohort included imaging studies of patients with and without a history of hip arthroplasty. We retrieved the Digital Imaging and Communications in Medicine (DICOM) images for the aforementioned studies and anonymized them by converting all studies into Neuroimaging Informatics Technology Initiative (NIfTI) files. Through a combination of programming pipelines and manual review, we searched for studies that met the following conditions: 1. captured the entire region of the pelvis (the area between the iliac crest superiorly and the lesser trochanter of the femurs inferiorly); 2. captured both hip joints; 3. had at least 50 axial slices; 4. had a slice thickness of 1.5mm or less; and 5. were axial images. This filtering process yielded 5,052 eligible studies that were then resampled to a pixel spacing of (1mm × 1mm × 1mm) and saved as our data pool for the training and validation of our pipeline. In cases where more than one series was available from a patient, we included all eligible series.

Following the collection of eligible CT studies, we leveraged the *DeepDRR* Python package to convert CT data into DRRs (20). Although conventional DRRs are only obtained using image processing of the underlying CT data, the DeepDRR method relies on a DL-based pipeline consisting of material decomposition of the CT data, ray-tracing forward projection, Rayleigh scatter estimation, and random noise injection to build the DRRs. To extract DRRs, we: 1) placed the CT in the isocentric (standard) position where the patient is head-first and supine (AP imaging); 2) obtained a baseline DRR ($DRR_0$) by AP projection; 3) rotated the CT using the $P_1 = [x_1, y_1, z_1]$ vector with the center of the volume as the anchor, where $\{x_1, y_1, z_1\}$ are randomly sampled integer values from $[-15°, +15°]$ and obtained $DRR_1$; 4) created a transformation vector $V = [a, b, c]$, where $\{a, b, c\}$ are sampled from $[-15°, +15°]$, and the absolute values of the angles in $P_2 = P_1 + V$ in each axis (x, y, z) were less than $15°$, i.e., $|P_{2x}| < 15°$ & $|P_{2y}| < 15°$ & $|P_{2z}| < 15°$; 5) rotated the CT to the new position $P_2$, and obtained $DRR_2$; 6) standardized all DRRs to a uniform size of 256×256 pixels, 7) saved the DRRs to disk after rescaling them to a pixel intensity range of 0 to 1, and 8) calculated and saved the coordinates of vectors that could convert could convert each DRR into other rotations of that DRR (i.e., $DRR_1{\rightarrow}DRR_2$, $DRR_2{\rightarrow}DRR_1$, $DRR_1{\rightarrow}DRR_0$, $DRR_2{\rightarrow}DRR_0$, $DRR_0{\rightarrow}DRR_1$, and $DRR_0{\rightarrow}DRR_2$). Overall, we created 512 DRRs for each of the 5,052 available CT volumes, resulting in over 2.5 million DRRs saved on disk. This produced more than 15 million pairs of input and output images to serve as the ground truth for training the DL model.

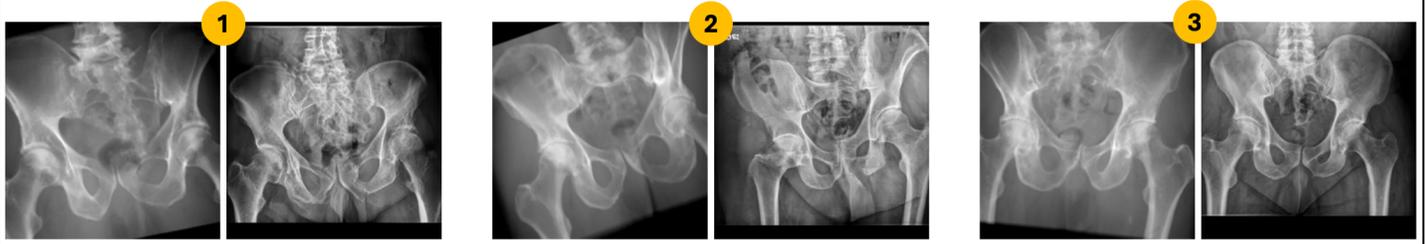
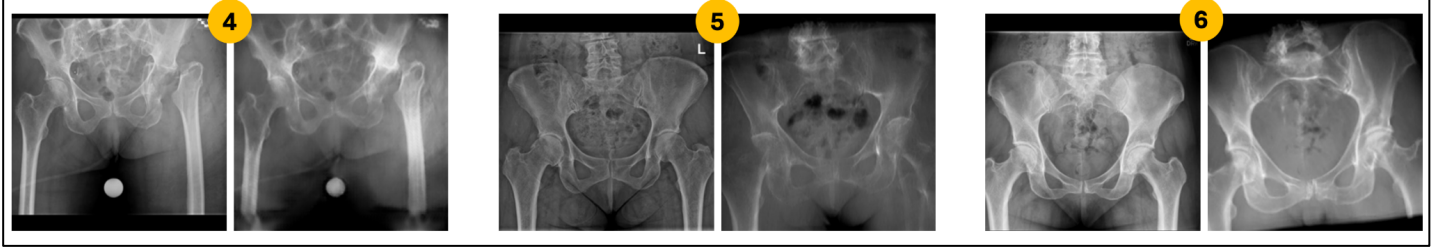

**Figure 4**: Examples of Cycle-GAN output for style transformation from real DRRs to X-Rays (1, 2, 3), and the reverse (4, 5, 6). Note the content alterations from real DRRs to synthetic X-Rays during the style transfer process. Abbreviations: *DRR*, Digitally Reconstructed Radiographs; *XR*, X-Ray.

Last but not least, we divided our data into seven random folds, grouping the DRRs at the patient level and stratifying the splits based on the acquisition date, number of slices, and slice thickness of the original CT data. Folds 1 through 5 were designated for training the DL models, Fold 6 was used for validation and hyperparameter tuning, and Fold 7 was reserved as our internal test set.

### 3.2 DRR to X-Ray style transfer.

Similar to previous studies, we initially attempted to train a DL-based style transfer model to convert the styles of DRRs in our training data to that of conventional radiographs before training the DDPM. To achieve this, we compiled a set of 250,000 DRRs (pooled from the CT data described above) as well as 250,000 AP radiographs from our institutional Total Hip Arthroplasty Radiography Registry (21). We then proceeded to train a Cycle-GAN, following the methodology introduced in (11), at a constant learning rate of 3e-4 until it began to overfit our dataset. However, the performance of the Cycle-GAN proved to be inconsistent during inference. In some cases, the style transfer occurred without noticeable issues, yet in others, the content of the original image was significantly altered during the style transfer process (**Figure 4**). We assume that at least part of this suboptimal performance could be attributed to the fact that the anatomy present in some extremely rotated DRRs is significantly different from that in the AP radiographs. The GAN architecture, which is well-known for learning entanglements, may not successfully apply the style of non-rotated AP radiographs to rotated DRRs without inevitably altering the content of the DRRs simultaneously. Notably, we also tried enhancing our training pool with non-AP radiographs (cross-table lateral and oblique views) to address this problem, however, that did not improve the Cycle-GAN performance.

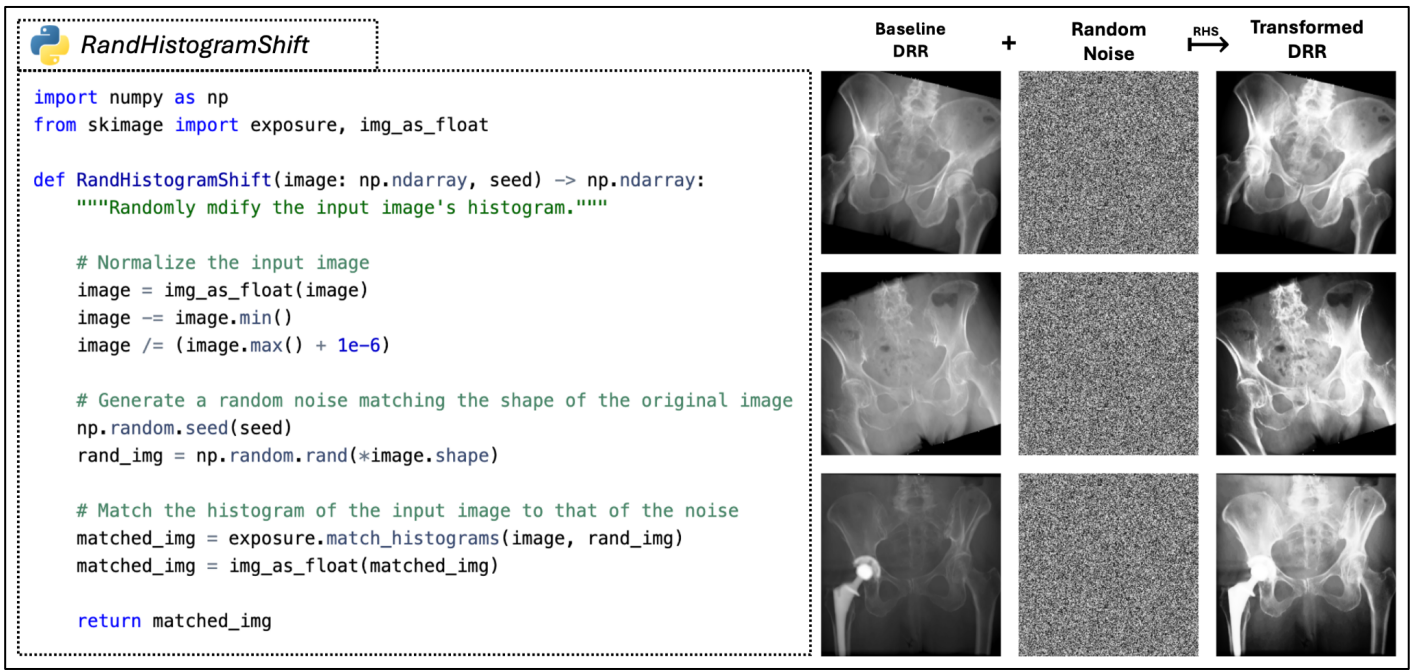

**Figure 5**. The Python code for the RandHistogramShift transformation (right) and examples of its application to random DRRs (left). Abbreviations: *DRR*, Digitally Reconstructed Radiographs, *RHS*, RandHistogramShift.

As a result of the Cycle-GAN performance, we replaced it with the inclusion of the ***RandHistogramShift*** transformation in our data preprocessing pipeline during the training phase. The logic behind the RandHistogramShift transfer is straightforward and effective. We hypothesized that the primary distinction between a DRR and a radiograph lies in their pixel intensity distributions, affecting their overall brightness and contrast. Therefore, by making the DDPM agnostic to the pixel intensity distribution of the input images, it should, in theory, become capable of generalizing to any type of radiography data (e.g., X-rays, DRRs, or fluoroscopy shots), despite being solely trained on DRRs. Our approach implementation is shown in **Figure 5**. During training, we generated a random image filled with stochastic noise and then applied image processing techniques to align the histogram of each pair of input and output DRRs with that of the noisy image. Given that the noisy image is randomly generated at each training step, the DDPM is pushed to disregard the stochastic pixel intensity distributions of the ground truth DRR pairs, focusing instead on their content to learn the 3D rotations. The outputs of applying the RandHistogramShift on DRRs and radiographs are shown in **Figure 5** and **Supplements**, respectively.

## 3.3 Model

Following the collection of DRRs, we trained a conditional DDPM that received an input DRR as well as three integer values representing the desired amount of transformations on the x, y, and z axes, respectively, and generated a transformed DRR as its output (**Figure 6**). To train the DL model, we established a forward diffusion process consisting of 1,000 steps, adding Gaussian noise to the original images and encouraging the model to learn the reverse diffusion process. This reverse diffusion process aims to approximate the noise added between two successive steps, thus reconstructing the less noisy image from the noisier version at each step (22). The DL model used for the reverse diffusion process features a U-Net-like architecture with several advantages over the vanilla U-Net, including multiple residual blocks in each layer, self-attention modules for improved image quality, and an embedded timestep number that dictates the intensity of the added noise (10).

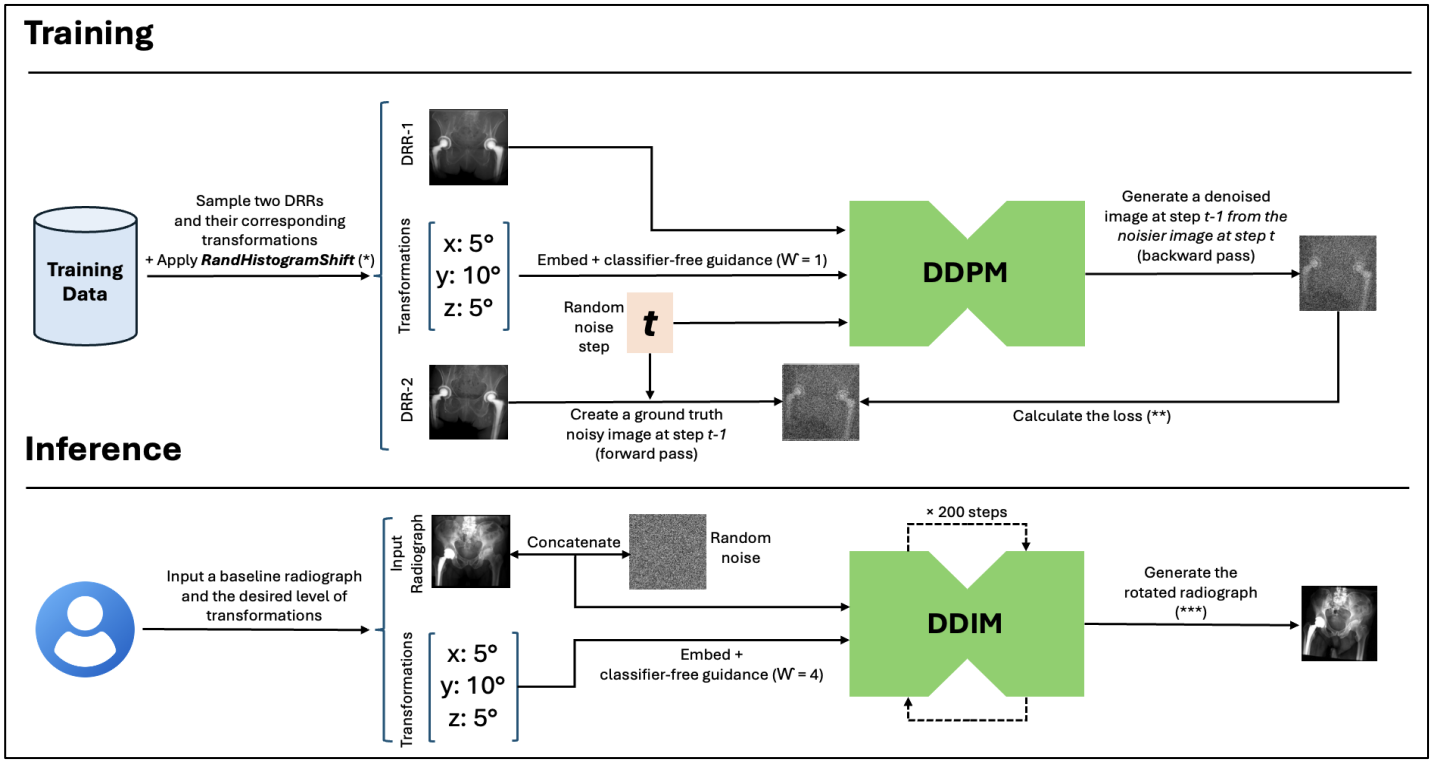

**Figure 6**. Training and inference pipelines for our proposed methodology. (*) Both DRRs underwent the RandHistogramShift transformation (with the same random noise) prior to input into the DDPM; this transformation is omitted in the figure for brevity. (**) For actual training, "v-prediction" was employed. Abbreviations: *DRR*, Digitally Reconstructed Radiograph; *DDPM*, Denoising Diffusion Probabilistic Model; *DDIM*, Denoising Diffusion Implicit Model. (***) The histogram of the generated radiograph will also be matched to that of the input radiograph (not shown in the figure).

Every DDPM requires an input image consisting of random Gaussian noise to gradually denoise it into a meaningful image through the reverse diffusion process. However, conditional DDPMs require additional signals, either in the form of imaging data or linear vectors, to generate the target image. In our case, we employed both strategies to provide the DDPM with everything it needed to generate a target DRR. First, we concatenated the original DRR with the noisy image as a second channel and then created a linear vector consisting of the three desired transformation values to guide the DDPM on the required changes to the original DRR. Conditioning the DDPM on this vector was accomplished using classifier-free guidance (23), a technique that obviates the need for loading an additional network to steer the DDPM generations toward certain classes. We have shared the detailed configurations of our DDPM in the **Supplements**.

## 3.4 Training

We trained the DDPM using the *Mediffusion* Python framework (24), with a constant learning rate of 0.0001, and utilized v-prediction as the training objective for our diffusion model to ensure correct image intensity during inference (25). We continued the training until an identical network, utilizing the exponential moving average of the weights,

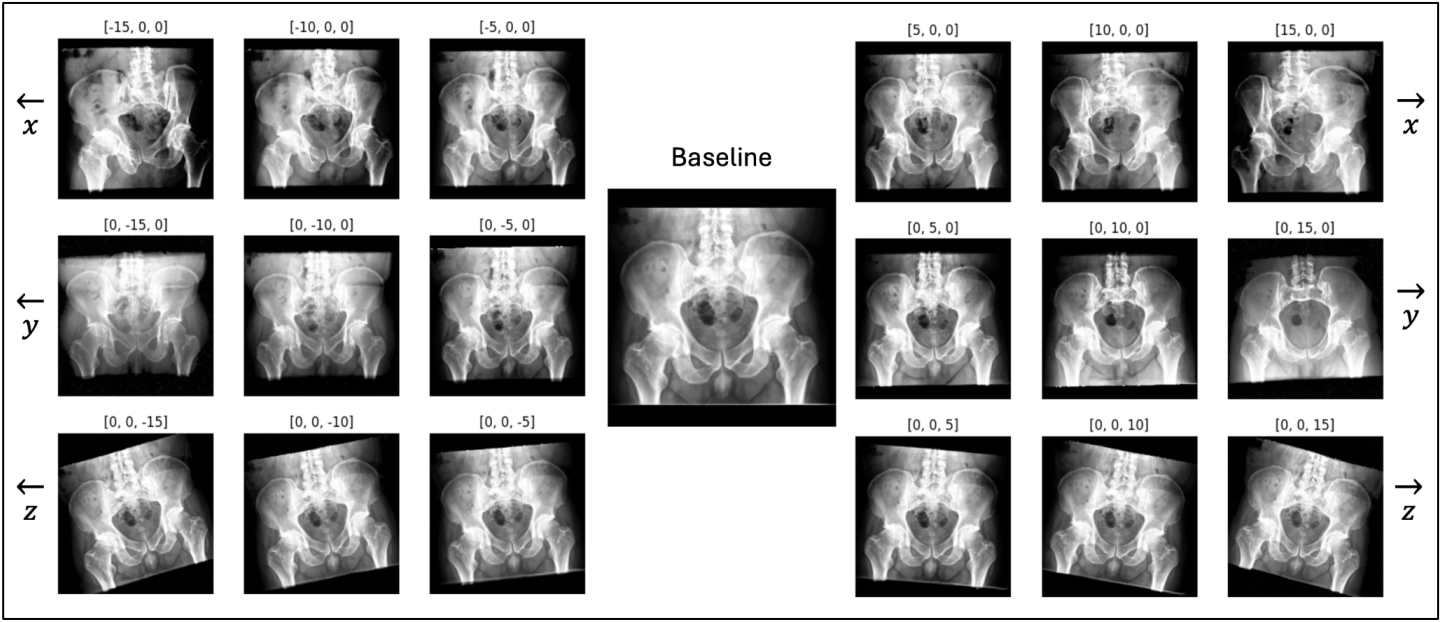

**Figure 7**. Example rotations of a preoperative radiograph in x, y, and z axes.

with a decay factor of 0.9999, from the actively trained network, overfit the validation set. We scaled all images to a pixel value range of [-1, 1] before feeding them into the DDPM model. In addition, we applied three specific transformations to the training set:

First, we added two rows of zero-valued pixels to the top and bottom margins of the input and output images. Extracted DRRs from the CT data do not have margins of zero pixels; however, the addition of such margins (i.e., padding) is inevitable for creating square-shaped radiographs during inference time. Including this transformation during training helps the DDPM not to regard the padded radiographs as outliers during inference. The number of images subjected to this transformation and the length of the added bars were determined randomly. Second, we randomly added rectangles of stochastic size to the input DRRs. This transformation aimed to mimic the appearance of radiographs with covered protected health information using black rectangles using an already established model (26). Last but not least, we leveraged the *RandHistogramShift* transformation during training to instruct the DDPM on transforming both DRRs and conventional radiographs during inference time.

## 4    Performance

We envisioned two main inference modes for our model: A. Use it in a single-shot or few-shot manner, where the user inputs a set of transformation angles in addition to a baseline radiograph and expects the model to rotate it according to the provided angles. B. Use it to generate an entire range of rotations in one or more axes, allowing the output frames to create a 30 frame-per-second (FPS) video clip of the anatomical volume being rotated in 3D. For both modes, we employed the Denoising Diffusion Implicit Model (DDIM) protocol with 200 inference steps and a conditioning scale of 4 for classifier-free guidance (27). The noise input to the model is randomly generated for each baseline image but is kept constant when generating multiple unseen views from a single baseline image (e.g., in mode B or in few-shot mode A). Keeping the noise vector constant and using a deterministic approach for running inference with DDIM ensures that multiple generations from a single baseline radiograph are more consistent with each other.

Mode A is cost-effective and fast. On a single A100 NVIDIA Tensor Core Graphics Processing Unit (GPU), it takes 10 to 12 seconds to generate a novel view. This mode is therefore suitable for instances where the user needs to visualize a more desirable view of a baseline radiograph, estimating how much and in which axes the radiograph should be rotated. Conversely, Mode B is more expensive and time-consuming, requiring 300 generations to cover a range of -15° to +15° in each axis: 30 images generated at each angle, multiplied by ten interpolated generations for each 1° of angle shift. Using batch inference, generating all the frames for a single axis will, on average, take 7 minutes.

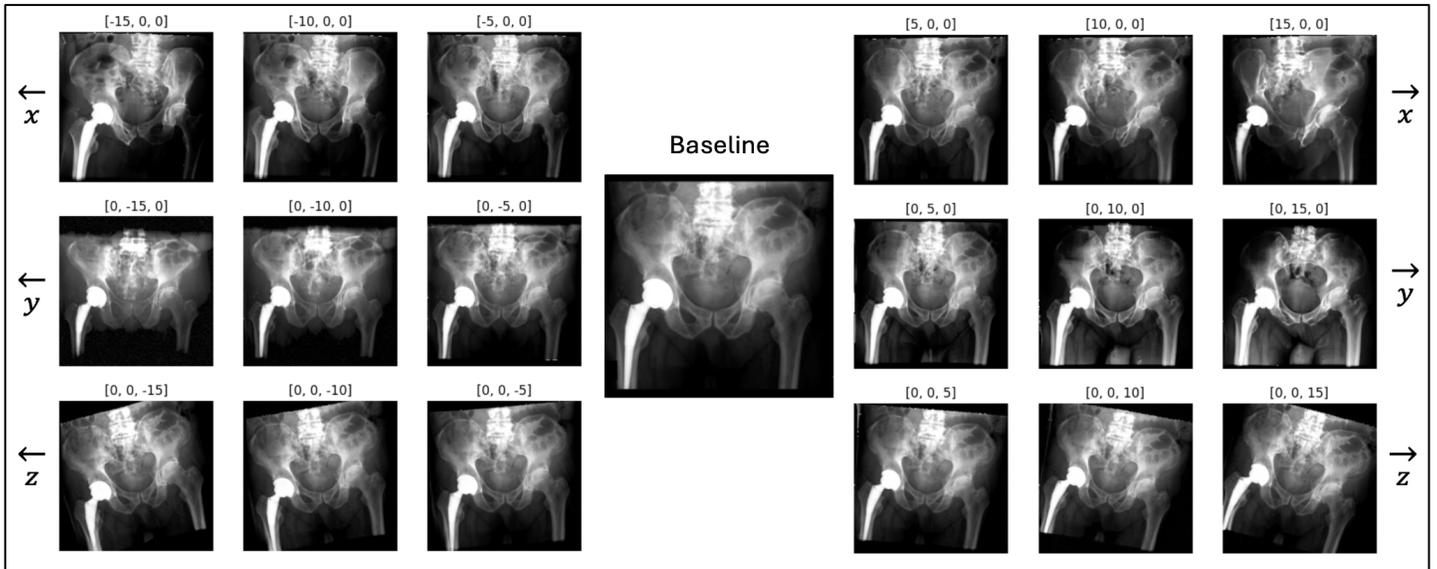

**Figure 8**. Example rotations of a postoperative radiograph in x, y, and z axes.

Therefore, Mode B is more suited for scenarios where the user prefers to visualize an entire range of rotations for observing certain anatomical (or pathological) features or to choose specific views from the many available.

**Figures 7 and 8** show the performance of our model in mode A for a random preoperative radiograph and a random postoperative radiograph, respectively. More examples of our model's performance, including its performance on input DRRs and also simultaneous rotations in two or three axes are provided in our Supplements. Similarly, an online demo of our tool, demonstrating the real-time performance of our model on a limited set of input imaging, is available through **HuggingFace Spaces**[1]. You can also visit **our website**[2] to watch several video clips showing how the model was able to generate over 300 consistent frames for rotating an input radiograph in Mode B.

Upon initial visual inspection, the DDPM displays acceptable and consistent overall performance in both modes. The generated radiographs match the anatomy of the baseline input radiograph, and the 3D rotation of various anatomical structures appears reasonable and consistent to both reviewing orthopedic surgeons and radiologists. As expected, we observed that our model had learned all possible rotations up to ±15° in the x, y, or z axes. However, the model did not learn to extrapolate to any rotations beyond ±15°, such as 45° or 90°. We hypothesize that this limitation could be overcome by training the model on a diverse set of input DRRs that include a more comprehensive range of ground truth rotations. Nevertheless, we acknowledge that a rigorous evaluation of our model's performance requires a systematic approach with expert inspection and cadaveric studies. We reserve such detailed discussions for our future works.

---

[1] https://pouriarouzrokh.github.io/RadRotator
[2] https://huggingface.co/spaces/Pouriarouzrokh/RadRotator

# 5     Conclusions

DL-based approaches have already been attempted to generate 3D volumes from two or more 2D medical imaging inputs (5,6). In this report, we further advanced these previous experiments by introducing a DDPM-based methodology that can rotate an input radiograph or DRR along the x, y, and/or z axes. Our method could be used to rotate an input 2D image in 3D (akin to changing the position of the X-ray source in actual radiography) or to demonstrate an entire anatomic volume in 3D by generating hundreds of consistently rotated frames from a single 2D baseline image. Similar to previous studies, we relied on CT volumes to acquire DRR training data for our DL model. However, we addressed two significant bottlenecks encountered in previous studies: 1. We utilized diffusion models instead of GANs to achieve higher mode coverage and output image quality, with the only sacrifice being the inference time, which is often less critical in medical applications; and 2. We demonstrated that the unreliable use of style transfer DL models, such as Cycle-GAN, could be replaced with a training transformation named RandHistogramShift. This transformation enables the reliable training of a DL model on input DRRs and the use of the exact same model for running inference on radiographs.

# 6     References

1.  Iyengar KP, Jun Ngo VQ, Jain VK, Ahuja N, Hakim Z, Sangani C. What does the orthopaedic surgeon want in the radiology report? Journal of Clinical Orthopaedics and Trauma. 2021;21:101530.

2.  Goodwin ML, Buchowski JM, Sciubba DM. Why X-rays? The importance of radiographs in spine surgery. Spine J. 2022;22(11):1759–1767.

3.  Carreira DS, Emmons BR. The Reliability of Commonly Used Radiographic Parameters in the Evaluation of the Pre-Arthritic Hip: A Systematic Review. JBJS Rev. 2019;7(2):e3.

4.  Thippeswamy PB, Nedunchelian M, Rajasekaran RB, Riley D, Khatkar H, Rajasekaran S. Updates in postoperative imaging modalities following musculoskeletal surgery. J Clin Orthop Trauma. 2021;22:101616.

5.  Ying, Guo, Ma, Wu, Weng. X2CT-GAN: reconstructing CT from biplanar X-rays with generative adversarial networks. Proc Estonian Acad Sci Biol Ecol. http://openaccess.thecvf.com/content_CVPR_2019/html/Ying_X2CT-GAN_Reconstructing_CT_From_Biplanar_X-Rays_With_Generative_Adversarial_Networks_CVPR_2019_paper.html.

6.  Ge R, He Y, Xia C, et al. X-CTRSNet: 3D cervical vertebra CT reconstruction and segmentation directly from 2D X-ray images. Knowledge-Based Systems. 2022;236:107680.

7.  Hampali S. 3D shape reconstruction of knee bones from low radiation X-ray images using deep learning. University of Waterloo; 2021. https://uwspace.uwaterloo.ca/handle/10012/17076. Accessed January 10, 2024.

8.  Shi Y, Wang P, Ye J, Long M, Li K, Yang X. MVDream: Multi-view Diffusion for 3D Generation. arXiv [cs.CV]. 2023. http://arxiv.org/abs/2308.16512.

9.  Ho J, Jain A, Abbeel P. Denoising Diffusion Probabilistic Models. arXiv [cs.LG]. 2020. http://arxiv.org/abs/2006.11239.

10. Dhariwal P, Nichol A. Diffusion models beat GANs on image synthesis. arXiv [cs.LG]. 2021. http://arxiv.org/abs/2105.05233.

11. Zhu J-Y, Park T, Isola P, Efros AA. Unpaired image-to-image translation using cycle-consistent adversarial networks. 2017 IEEE International Conference on Computer Vision (ICCV). IEEE; 2017. p. 2223–2232.

12. Xiao Z, Kreis K, Vahdat A. Tackling the Generative Learning Trilemma with Denoising Diffusion GANs. arXiv [cs.LG]. 2021. http://arxiv.org/abs/2112.07804.

13. Milickovic N, Baltast D, Giannouli S, Lahanas M, Zamboglou N. CT imaging based digitally reconstructed radiographs and their application in brachytherapy. Phys Med Biol. 2000;45(10):2787–2800.

14. Moore CS, Avery G, Balcam S, et al. Use of a digitally reconstructed radiograph-based computer simulation for the optimisation of chest radiographic techniques for computed radiography imaging systems. Br J Radiol. 2012;85(1017):e630–e639.

15. Li X, Zhou L-H, Zhen X, Lu W-T. The Generation of Digitally Reconstructed Radiographs with Six Parameters. 2010 4th International Conference on Bioinformatics and Biomedical Engineering. IEEE; 2010. p. 1–4.

16. Yang C-J, Lin C-L, Wang C-K, et al. Generative Adversarial Network (GAN) for Automatic Reconstruction of the 3D Spine Structure by Using Simulated Bi-Planar X-ray Images. Diagnostics (Basel). 2022;12(5). doi: 10.3390/diagnostics12051121.

17. Gao C, Killeen BD, Hu Y, et al. Synthetic data accelerates the development of generalizable learning-based algorithms for X-ray image analysis. Nat Mach Intell. 2023;5(3):294–308.

18. Loÿen E, Dasnoy-Sumell D, Macq B. Patient-specific three-dimensional image reconstruction from a single X-ray projection using a convolutional neural network for on-line radiotherapy applications. Phys Imaging Radiat Oncol. 2023;26:100444.

19. Hoyez H, Schockaert C, Rambach J, Mirbach B, Stricker D. Unsupervised Image-to-Image Translation: A Review. Sensors . 2022;22(21). doi: 10.3390/s22218540.

20. Unberath M, Zaech J-N, Lee SC, et al. DeepDRR – A Catalyst for Machine Learning in Fluoroscopy-Guided Procedures. Medical Image Computing and Computer Assisted Intervention – MICCAI 2018. Springer International Publishing; 2018. p. 98–106.

21. Rouzrokh P, Khosravi B, Johnson QJ, et al. Applying deep learning to establish a total hip arthroplasty radiography registry: a stepwise approach. JBJS. LWW; 2022;104(18):1649–1658.

22. Nichol A, Dhariwal P. Improved denoising diffusion probabilistic models. arXiv [cs.LG]. 2021. http://arxiv.org/abs/2102.09672.

23. Ho J, Salimans T. Classifier-Free Diffusion Guidance. arXiv [cs.LG]. 2022. http://arxiv.org/abs/2207.12598.

24. Khosravi B, Rouzrokh P, Mickley JP, et al. Few-shot Biomedical Image Segmentation using Diffusion Models: Beyond Image Generation. Comput Methods Programs Biomed. Elsevier; 2023;107832.

25. Salimans T, Ho J. Progressive Distillation for Fast Sampling of Diffusion Models. arXiv [cs.LG]. 2022. http://arxiv.org/abs/2202.00512.

26. Khosravi B, Mickley JP, Rouzrokh P, et al. Anonymizing Radiographs Using an Object Detection Deep Learning Algorithm. Radiology: Artificial Intelligence. Radiological Society of North America; 2023;e230085.

27. Song J, Meng C, Ermon S. Denoising Diffusion Implicit Models. arXiv [cs.LG]. 2020. http://arxiv.org/abs/2010.02502.

# 7 Supplements

## 7.1 Additional Performance Examples

Additional examples to showcase our model's capability to randomly rotate input imaging data. The notation [a, b, c] represents the random degree of rotation along the x, y, and z axes, respectively, where a, b, and c are in [-15°, 15°].

### 7.1.1 Performance on input radiographs:

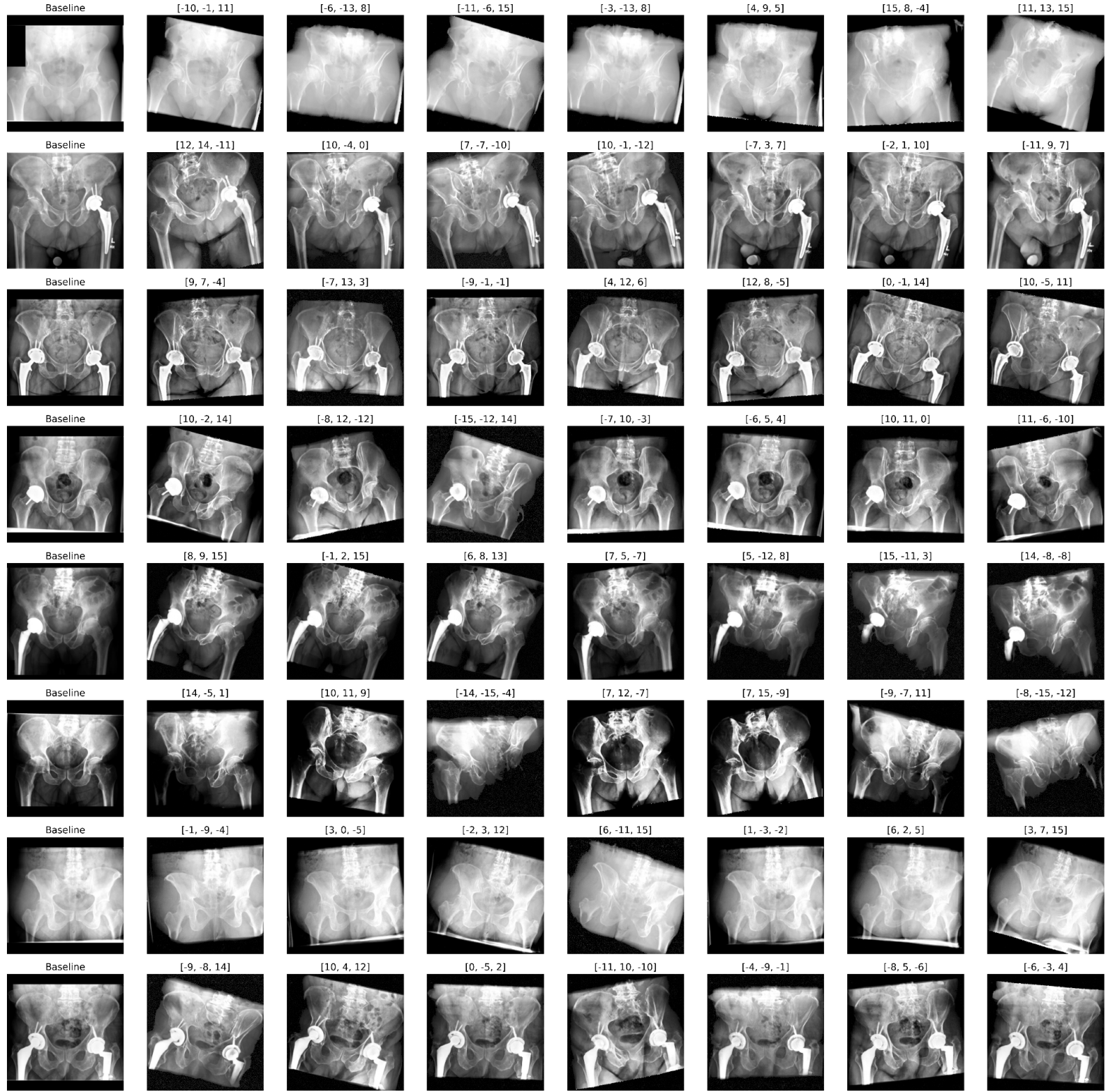

### 7.1.2 Performance on input Digitally Reconstructed Radiographs (DRRs):

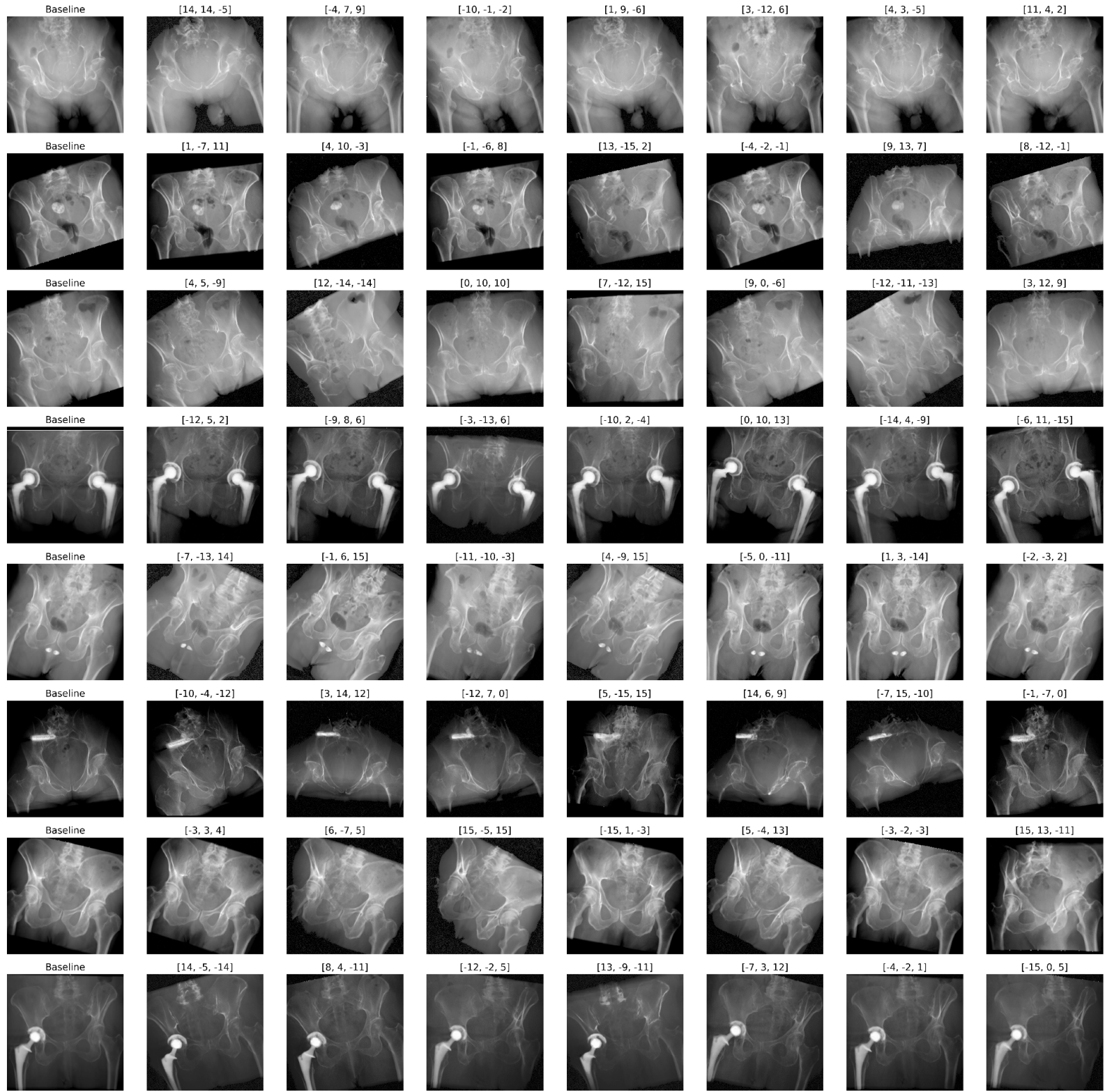

## 7.2 Configurations used for the training and validation of our DDPM

These configurations are based on the Mediffusion package. For a detailed introduction to these settings, please visit https://github.com/BardiaKh/Mediffusion.

| Parameter | Value | Parameter | Value |
|---|---|---|---|
| Diffusion | | input_size | 256 |
| timesteps | 1000 | dims | 2 |
| schedule_name | cosine | attention_resolutions | [8, 16, 32] |
| enforce_zero_terminal_snr | true | channel_mult | [1, 1, 2, 2, 4, 4] |
| schedule_params: | | dropout | 0.0 |
| - beta_start | 0.0001 | in_channels | 2 |
| - beta_end | 0.02 | out_channels | 2 |
| - cosine_s | 0.008 | model_channels | 128 |
| timestep_respacing | null | num_head_channels | -1 |
| mean_type | VELOCITY | num_heads | 4 |
| var_type | LEARNED_RANGE | num_heads_upsample | -1 |
| loss_type | Mean Squared Error | num_res_blocks | [2, 2, 2, 2, 2, 2] |
| Optimizer | | resblock_updown | false |
| Learning Rate | 1e-4 | use_checkpoint | false |
| Type | Lion | use_new_attention_order | false |
| Validation | | use_scale_shift_norm | true |
| classifier_cond_scale | 4 | scale_skip_connection | false |
| protocol | DDPM | Conditions | |
| log_original | true | num_classes | 4 |
| log_concat | true | concat_channels | 1 |
| cls_log_indices | [0, 1, 2, 3] | guidance_drop_prob | 0.1 |
| Model | | missing_class_value | null |

## 7.3 RandHistogramShift Transformation Applied to X-ray Radiographs.

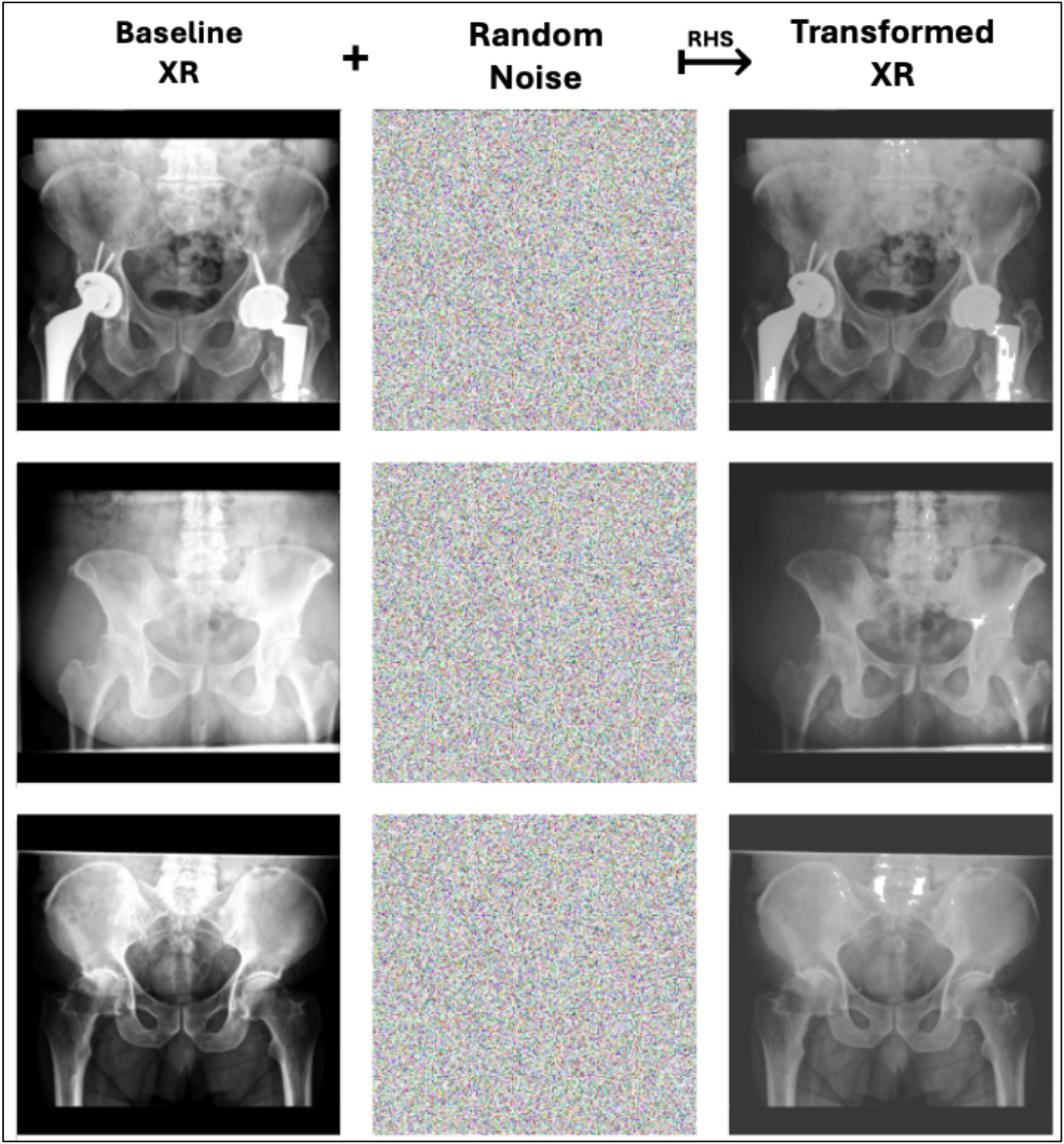